\documentclass[twocolumn,preprintnumbers,aps,showpacs,superscriptaddress,prl]{revtex4-1}

\usepackage{palatino}
\usepackage[latin1]{inputenc}
\usepackage{epsf}
\usepackage{amsmath,amssymb}
\usepackage{latexsym}
\usepackage{calc}
\usepackage{color}
\usepackage{shadow}
\usepackage{epsfig}
\usepackage{float}

\usepackage[pdftex, pdfstartview = {FitH}]{hyperref}

\newcommand{\ben}{\begin{equation}}
\newcommand{\een}{\end{equation}}
\newcommand{\bea}{\begin{eqnarray}}
\newcommand{\eea}{\end{eqnarray}}

\def\ext{_{\rm ext}}

\def\br{{\bf r}}
\def\bR{{\bf R}}

\def\dulR{{\underline{\underline{\bf R}}}}
\def\dulr{{\underline{\underline{\bf r}}}}

\begin{document}
  \title{Exact Electronic Potentials in Coupled Electron-Ion Dynamics}
\author{Yasumitsu Suzuki}
\affiliation{Max-Planck Institut f\"ur Mikrostrukturphysik, Weinberg 2,
D-06120 Halle, Germany}
\author{Ali Abedi}
\affiliation{Max-Planck Institut f\"ur Mikrostrukturphysik, Weinberg 2,
D-06120 Halle, Germany}
\author{Neepa T. Maitra} 
\affiliation{Department of Physics and Astronomy, Hunter College and the City University of New York, 
695 Park Avenue, New York, New York 10065, USA}
\author{Koichi Yamashita} 
\affiliation{Department of Chemical System Engineering, School of Engineering, The University of Tokyo, 
7-3-1 Hongo, Bunkyo-ku, Tokyo 113-8656, Japan}
\author{E.K.U. Gross}
\affiliation{Max-Planck Institut f\"ur Mikrostrukturphysik, Weinberg 2,
D-06120 Halle, Germany}

\date{\today}

\begin{abstract}
We develop a novel approach to the coupled motion of electrons and ions that focuses on the dynamics
of the electronic subsystem. Usually the description of electron dynamics involves an electronic 
Schr\"odinger equation where the nuclear degrees of freedom appear as parameters or as classical trajectories.
Here we derive the exact Schr\"odinger equation for the subsystem of electrons,
staying within a full quantum treatment of the nuclei. This exact Schr\"odinger
equation features a time-dependent potential energy surface for electrons (e-TDPES).
We demonstrate that this exact e-TDPES differs significantly from the electrostatic potential produced by classical or quantum nuclei. 
\end{abstract}

\maketitle
The theoretical description of electronic motion in the time domain is among the biggest challenges in 
theoretical physics.
A variety of tools has been developed to tackle this problem,
among them 
the Kadanoff-Baym approach~\cite{SR,*marini},
time-dependent density functional theory~\cite{RG,*carsten,*BSIN},
the hierarchical equations of motion approach~\cite{jin,*chen}
as well as
the multiconfiguration time-dependent Hartree-Fock approach~\cite{scrinzi,*kato,*WWCTR,*irene}.
From the point of view of electronic dynamics 
all these approaches are formally exact as long as the nuclei are considered clamped.
However, some of the most fascinating phenomena result from the coupling of 
electronic and nuclear motion, e.g., 
photovoltaics
~\cite{RFSR13,*DP07}, processes in vision~\cite{TTR07,*Polli}, photosynthesis~\cite{TMF11}, molecular electronics~\cite{HBFTM04,*VSC,*Ratner},
and strong field processes~\cite{CREI,*Esa,*Lein}. 
To properly capture
electron dynamics in these phenomena, it is essential to account for electron-nuclear (e-n) coupling.

In principle, the e-n dynamics is described by the
complete time-dependent Schr\"odinger equation (TDSE)
\ben
\label{eq:tdse}
 \hat{H} \Psi(\dulr,\dulR,t)= i\partial_t\Psi(\dulr,\dulR,t),
\een
with Hamiltonian
\begin{equation}
 \hat{H} = \hat{T}_{n}(\dulR)+ \hat{V}^{n}_{ext}(\dulR,t) +\hat{H}_{\rm BO}(\dulr,\dulR)
 +\hat{v}^{e}_{ext}(\dulr,t),
\label{eq:completeH}
\end{equation} 
where $\hat{H}_{\rm BO}(\dulr,\dulR)$ is the traditional Born-Oppenheimer
(BO) electronic Hamiltonian,
\begin{equation}
 \hat{H}_{\rm BO} = \hat{T}_{e}(\dulr)+ \hat{W}_{ee}(\dulr) +\hat{W}_{en}(\dulr,\dulR)+\hat{W}_{nn}(\dulR).
\label{eq:HBO}
\end{equation} 
Here $\hat{T}_{n}=-\sum_{\alpha=1}^{N_n}\frac{\nabla^2_\alpha}{2M_\alpha}$ and 
 $\hat{T}_{e}=-\sum_{j=1}^{N_e}\frac{\nabla^2_j}{2m}$ are the nuclear and electronic kinetic energy operators,
 $\hat{W}_{ee}$, $\hat{W}_{en}$ and $\hat{W}_{nn}$ are the electron-electron, e-n
and nuclear-nuclear interaction, and $\hat{V}^{n}_{ext}(\dulR,t)$ and $\hat{v}^{e}_{ext}(\dulr,t)$ are 
time-dependent (TD) external potentials acting on the nuclei and electrons, respectively. 
Throughout this paper $\dulR$ and $\dulr$ collectively represent the nuclear and electronic coordinates respectively
and $\hbar=1$. 

A full numerical solution of the complete e-n TDSE, Eq. (\ref{eq:tdse}), is extremely hard to achieve and has been obtained 
only for small systems with very few degrees of freedom, such as H$_2^+$~\cite{Bandrauk}. For larger systems, an 
efficient and widely used approximation is the mixed quantum-classical description where the electrons are propagated 
quantum mechanically according to the TDSE
 \ben
 \label{eq:trad_e_td}
 \Bigl( \hat{T}_{e}(\dulr) +\hat{W}_{ee}(\dulr) + V(\dulr,t) + \hat{v}^{e}_{ext}(\dulr ,t)
 \Bigr)\Phi(\dulr,t)=i\partial_t \Phi(\dulr,t),       
 \een
 which is coupled to the classical nuclear trajectories, ${\bf R}_\alpha (t)$, determined by Ehrenfest or surface-hopping 
 algorithms~\cite{tully}.
 The potential $V(\dulr,t)$ felt by the electrons is then given by the classical expression
\ben
V_{class}(\dulr,t)=W_{en}(\dulr, \dulR(t)) = -\sum_{j=1}^{N_e}\sum_{\alpha=1}^{N_n} 
\frac{eZ_\alpha}{\vert \br_j - \bR_\alpha(t)\vert},
\label{eq:Vconv}
\een
where $\dulR(t)$ denotes the set of classical nuclear trajectories, ${\bf R}_\alpha (t)$.
A better approximation to the potential $V(\dulr,t)$ felt by the electrons is the electrostatic or Hartree expression~\cite{tully}:
\ben
V_{\rm Hartree}(\dulr,t)  = -eZ_\alpha\sum_{j=1}^{N_e}\sum_{\alpha=1}^{N_n} \int d\dulR\frac{\vert\chi(\dulR,t)\vert^2}{\vert \br_j - \bR_\alpha \vert} \;
\label{eq:VTDH}
\een
where $\chi(\dulR,t)$ represents a nuclear many-body wavefunction obtained, e.g., from nuclear wave packet dynamics. Clearly, Eq.~(\ref{eq:VTDH}) 
reduces to the classical expression~(\ref{eq:Vconv}) in the limit of very narrow wave packets centered around the 
classical trajectories 
$\dulR(t)$. The Hartree expression~(\ref{eq:VTDH}) incorporates the nuclear charge distribution,
 but the potential is still approximate as it neglects e-n correlations. 

In this paper we address the question whether the potential $V(\dulr,t)$ in the purely electronic many-body TDSE, Eq.~(\ref{eq:trad_e_td}), can 
be chosen such that the resulting electronic wavefunction $\Phi(\dulr,t)$ becomes {\it exact}. By exact we mean that $\Phi(\dulr,t)$ reproduces the 
true electronic $N_e$-body density and the true $N_e$-body current density that would be obtained from the full e-n wavefunction $\Psi(\dulr,\dulR,t)$ 
of Eq.~(\ref{eq:tdse}). We shall demonstrate that the answer is yes provided we allow for a vector potential, ${\bf S}(\dulr,t)$,
 in the electronic TDSE, in addition to the scalar potential $V(\dulr,t)$. 
 We will analyse this potential for an exciting experiment, namely the 
laser-induced localization of the electron in the H$_2^+$ molecule~\cite{sansone,*HRB,*KSIV}. 
We find significant differences between this exact potential and both the
 classical-nuclei potential Eq.~(\ref{eq:Vconv}) and the Hartree potential Eq.~(\ref{eq:VTDH}).

Refs.~\cite{AMG,AMG2}  proved that  
the {\it exact} solution of the complete molecular TDSE Eq.~(\ref{eq:tdse}) can be written as a single product, 
\ben\label{eq:fact1} 
\Psi(\dulr,\dulR,t)=\Phi_{\dulR}(\dulr,t)\chi(\dulR,t),
\een
 of a nuclear wavefunction $\chi(\dulR,t)$, and an electronic wavefunction parametrized by the nuclear coordinates, $\Phi_{\dulR}(\dulr,t)$, 
 which satisfies the partial normalization condition (PNC) $\int d\dulr\vert\Phi_{\dulR}(\dulr,t)\vert^2=1$.
Here we instead consider the {\it reverse} factorization, 
\ben\label{eq:fact2} 
\Psi(\dulr,\dulR,t)=\chi_{\dulr}(\dulR,t)\Phi(\dulr,t)\;.
\een
It is straightforward to see that the formalism presented in Ref.~\cite{AMG} 
follows through simply with a switch of the role of electronic and nuclear coordinates. 
In particular, 

(i) The exact solution of the TDSE may be written as Eq.~(\ref{eq:fact2}), where $\chi_{\dulr}(\dulR,t)$ satisfies the PNC 
$\int d\dulR\vert\chi_{\dulr}(\dulR,t)\vert^2=1$. 

(ii) The nuclear wavefunction $\chi_{\dulr}(\dulR,t)$ satisfies
\ben
 \label{eq:exact_nucl_td}       
 \Bigl(\hat{H}_{n}(\dulR,\dulr,t)-\epsilon_{e}(\dulr,t)\Bigr)\chi_{\dulr}(\dulR,t)\\=i\partial_t \chi_{\dulr}(\dulR,t),
\een
with the nuclear Hamiltonian 
\ben
 \label{eq:n_ham_td}
 \begin{split}
  \hat{H}_{n}&(\dulR,\dulr,t) =   \hat{T}_n(\dulR) + \hat{W}_{ee}(\dulr) + \hat{W}_{en}(\dulr,\dulR)+ \hat{W}_{nn}(\dulR) \\
  &+\hat{v}\ext^e(\dulr,t)+ \hat{V}\ext^n(\dulR,t)  +\sum_{j=1}^{N_e}\frac{1}{m} \Big[\frac{(-i\nabla_j-{\bf S}_j(\dulr,t))^2}{2} \\
  &+ \Big(\frac{-i\nabla_j \Phi}{\Phi}+{\bf S}_j(\dulr,t)\Big)\left(-i\nabla_j-{\bf S}_j(\dulr,t)\right)\Big].
 \end{split}
\een
The electronic wavefunction $\Phi(\dulr,t)$ satisfies the TDSE:
\ben
 \label{eq:exact_e_td} 
  \Bigl(\sum_{j=1}^{N_e}\frac{1}{2m}(-i\nabla_j+{\bf S}_j(\dulr,t))^2 + \epsilon_{e}(\dulr,t)\Bigr)\Phi(\dulr,t)=i\partial_t \Phi(\dulr,t).            
\een
Here the exact TD potential energy surface for electrons (e-TDPES) $\epsilon_{e}(\dulr,t)$ and 
the exact electronic TD vector potential ${\bf S}_j(\dulr,t)$ are defined as
\ben
 \label{eq:exact_etdpes}
 \epsilon_{e}(\dulr,t) = \left\langle\chi_{\dulr}(t) \right\vert\hat{H}_{n}(\dulR,\dulr,t) - i \partial_t\left\vert
 \chi_{\dulr}(t)\right\rangle_\dulR 
\een
\ben
 \label{eq:exact_evect}
 {\bf S}_j(\dulr,t)=\left\langle\chi_{\dulr}(t)\right\vert\left.-i\nabla_j\chi_\dulr(t)\right\rangle_\dulR
\een
where $\langle ...|...|...\rangle_\dulR$ denotes an inner product over all nuclear variables only.

(iii) Eqs.~(\ref{eq:exact_nucl_td})-~(\ref{eq:exact_e_td}) are form-invariant under
the following gauge-like transformation
$\chi_{\dulr}(\dulR,t)\rightarrow\tilde{\chi}_{\dulr}(\dulR,t)=\exp(i\theta(\dulr,t))\chi_{\dulr}(\dulR,t)$, $\Phi(\dulr,t)\rightarrow\tilde
{\Phi}(\dulr,t)=\exp(-i\theta(\dulr,t))\Phi(\dulr,t)$, while the potentials transform as 
${\bf S}_j(\dulr,t)\rightarrow\tilde{\bf S}_j(\dulr,t)={\bf S}_j(\dulr,t)+\nabla_j
\theta(\dulr,t)$,
$\epsilon_{e}(\dulr,t)\rightarrow\tilde\epsilon_{e}(\dulr,t)=\epsilon_{e}(\dulr,t)+
\partial_t\theta(\dulr,t)$.
The wave functions $\chi_{\dulr}(\dulR,t)$ and $\Phi(\dulr,t)$ yielding a given solution, $\Psi(\dulr,\dulR,t)$,
of Eq.~(\ref{eq:tdse}) are unique up to this $(\dulr,t)$-dependent phase transformation.

(iv) The wave functions $\chi_{\dulr}(\dulR,t)$ and $\Phi(\dulr,t)$ are
 interpreted as nuclear and electronic wavefunctions:
$|\Phi(\dulr,t)|^{2}=\int |\Psi(\dulr,\dulR,t)|^{2}d\dulR$ is the probability density of finding
the electronic configuration $\dulr$ at time $t$, and 
$|\chi_{\dulr}(\dulR,t)|^{2}=|\Psi(\dulr,\dulR,t)|^{2}/|\Phi(\dulr,t)|^{2}$ is the conditional probability 
of finding the nuclei at $\dulR$, given that the electronic configuration is $\dulr$. The exact  electronic $N_e$-body 
current-density can be obtained from ${\rm Im}(\Phi^*\nabla_j\Phi)+|\Phi(\dulr,t)|^{2}{\bf S}_j$.

We can regard Eq.~(\ref{eq:exact_e_td}) as the {\it exact electronic TDSE}:
The time evolution of $\Phi(\dulr,t)$ is completely determined by the exact e-TDPES, 
$\epsilon_{e}(\dulr,t)$, and the vector potential ${\bf S}_j(\dulr,t)$.
Moreover, these potentials are {\it unique} up to within a gauge transformation (iii, above).
In other words, if one requires a purely electronic TDSE~(\ref{eq:exact_e_td}) with solution $\Phi(\dulr,t)$ to yield 
the true electron ($N_e$-body) density and current density of the full e-n problem, then the potentials appearing in this TDSE are 
(up to within a gauge transformation) uniquely given by Eqs.~(\ref{eq:exact_etdpes}) and~(\ref{eq:exact_evect}). 

A formalism in which the nuclear wavefunction is conditionally
dependent on the electronic coordinates, rather than the other way
around, may appear somewhat non-intuitive. However, in many non-adiabatic processes, the nuclear 
and electronic speeds are comparable, and, in some cases, such as highly excited Rydberg molecules, nuclei may even move faster 
than electrons~\cite{Rabani}. We shall show in the following that the present factorization is useful to interpret 
the dynamics of attosecond electron localization, and that it gives direct insight into how the e-n coupling affects non-adiabatic 
electron dynamics. For this purpose it is useful to rewrite the exact e-TDPES as
\ben
\epsilon_e(\dulr,t) = \epsilon^{\rm approx}_e(\dulr,t)+\Delta\epsilon_e(\dulr,t) 
\een 
where
\ben
\label{eq:eps_approx}
\begin{split}
\epsilon^{\rm approx}_e(\dulr,t)&=\left\langle\chi_\dulr(t) \right\vert 
 \hat{W}_{ee}(\dulr)+\hat{W}_{en}(\dulr,\dulR)+\hat{W}_{nn}(\dulR)\\
 &+\hat{v}^{e}_{ext}(\dulr,t)+\hat{V}^{n}_{ext}(\dulR,t)
 \left\vert
 \chi_\dulr(t)\right\rangle_\dulR 
\end{split}
\een
and
\ben
 \label{eq:Delta_eps}
 \begin{split}
\Delta\epsilon_e(\dulr,t) &= \left\langle\chi_{\dulr}(t) \right\vert \hat{T}_n(\dulR)
\left\vert \chi_{\dulr}(t)\right\rangle_\dulR+\left\langle\chi_{\dulr}(t)\right\vert - i \partial_t\left\vert \chi_{\dulr}(t)\right\rangle_\dulR\\
 &+\sum_{j=1}^{N_e}\frac{\left\langle \nabla_j \chi_{\dulr}(t) \vert \nabla_j \chi_{\dulr}(t)\right\rangle_\dulR }{2m}-\sum_{j=1}^{N_e}\frac{{\bf S}^2_j(\dulr,t)}{2m}. 
 \end{split} 
 \een 
If the nuclear density is approximated as a delta-function at $\dulR(t)$, then $\epsilon^{\rm approx}_e$ 
reduces to the electronic potential used in the traditional mixed quantum-classical approximations:
\ben
\label{trad_tdpes}
\begin{split}
\epsilon^{\rm trad}_{e}(\dulr,t)
=&\hat{W}_{ee}(\dulr) + 
\hat{W}_{en}(\dulr,\dulR(t))
 + \hat{W}_{nn}(\dulR(t))\\
  &+ \hat{v}^{e}_{ext}(\dulr ,t)+ \hat{V}^{n}_{ext}(\dulR(t)).
\end{split}
\een
This approximation not only neglects the width of the nuclear wavefunction  
but it also misses the contribution to the potential from $\Delta\epsilon_e(\dulr, t)$,
Eq.~(\ref{eq:Delta_eps}).  Methods that retain a quantum description
of the nuclei (e.g. TD Hartree ~\cite{tully}) approximate
Eq.~(\ref{eq:eps_approx}), although without the parametric dependence of the nuclear wavefunction 
 on $\dulr$, and still miss the contribution from
Eq.~(\ref{eq:Delta_eps}). In the following example, we will show the significance of the e-n correlation represented in the term $\Delta\epsilon_e$.

Among the many charge-transfer processes accompanying nuclear motion mentioned earlier, here we study attosecond
electron localization dynamics in the dissociation of the H$_2^+$
molecule achieved by 
time-delayed coherent ultrashort laser
pulses~\cite{sansone,*HRB,*KSIV}. In the experiment, first an ultraviolet
(UV) pulse excites H$_2^+$ to the dissociative $2p\sigma_u$ state while a
second time-delayed infrared (IR) pulse induces electron transfer
between the dissociating atoms.  This relatively recent  
technique has gathered increasing attention since it is expected
to eventually lead to the direct control of chemical reactions via the control of
electron dynamics.  Extensive theoretical studies have
led to progress in understanding the mechanism~\cite{sansone,*HRB,*KSIV}, and highlight the important role of e-n
correlated motion. Here we study 
the exact e-n coupling terms
 by computing the exact e-TDPES~Eq.~(\ref{eq:exact_etdpes}).
 
\begin{figure}[h]
 \centering
 \includegraphics*[width=1.0\columnwidth]{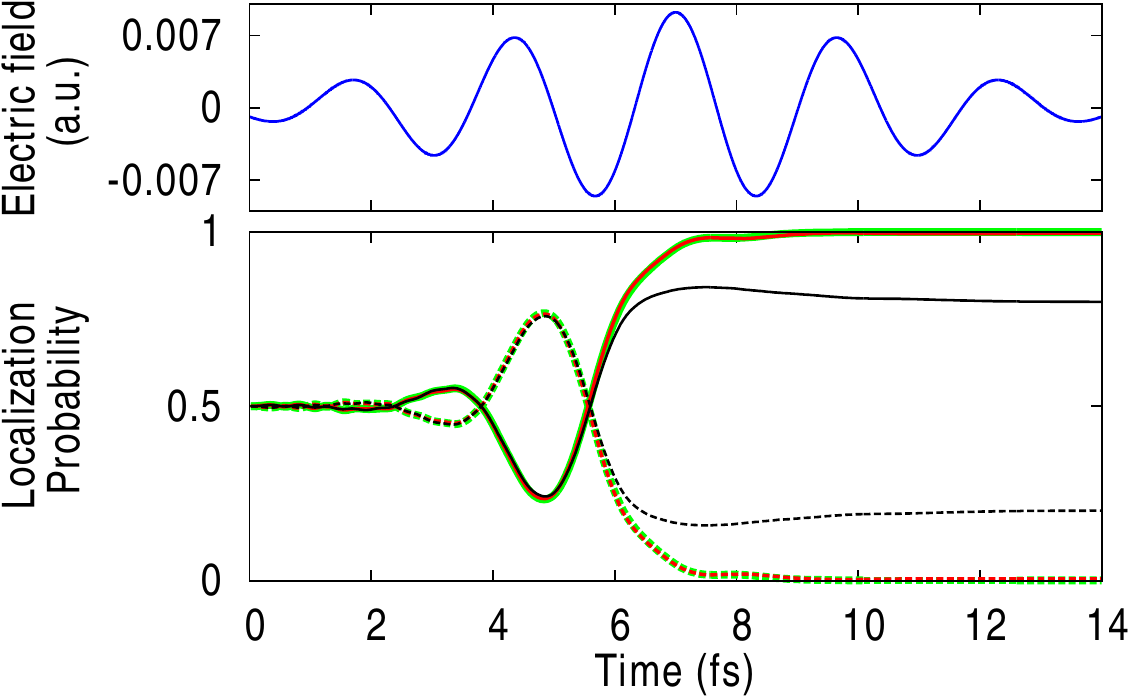}
 \caption{Electron localization probabilities along the negative (solid line) and 
 the positive z-axis (dashed line) as a function of time, obtained from exact dynamics (black), 
 dynamics on the traditional potential $\epsilon^{\rm trad}_{e}$
 evaluated at the exact mean nuclear position (red),
 and  dynamics on the approximate potential $\epsilon^{\rm approx}_{e}$ (green). The field is shown in the top panel.}
 \label{fig:Fig1}
\end{figure}

We consider a one-dimensional H$_2^+$ model, starting the dynamics after the excitation by the UV pulse: 
the wavepacket starts at $t = 0$ on the first excited state ($2p\sigma_u$ state) of H$_2^+$ as a Frank-Condon projection 
of the wavefunction of the ground state, and then is exposed to the  IR laser pulse.
The Hamiltonian is given by Eq.~(\ref{eq:completeH}), with $\dulR \to R$, the internuclear distance, and $\dulr \to z$, 
the electronic coordinate as measured from the nuclear center-of 
mass~\footnote{In this model system the electronic wavefunction $\Phi(z,t)$ and density 
$|\Phi(z,t)|^2$ are defined with respect to a coordinate frame attached to the nuclear framework so that they are 
characteristic for the internal 
properties of the system~\cite{KvG,*KG}.}. 
The kinetic energy terms are $\hat{T}_n(R) = -\frac{1}{2\mu_n}\frac{\partial^2}{\partial R^2}$ and, 
$\hat{T}_e(z) = -\frac{1}{2\mu_e}\frac{\partial^2}{\partial z^2}$, respectively, where the reduced mass of the nuclei is given by $\mu_n=M_{\rm H}/2$, 
and reduced electronic mass is given by $\mu_e=\frac{2M_{\rm H}}{2M_{\rm H}+1}$ ($M_{\rm H}$ is the proton mass).
The interactions are soft-Coulomb: $\hat{W}_{nn}(R) = \frac{1}{\sqrt{0.03+R^2}}$,
and $\hat{W}_{en}(z,R) = -\frac{1}{\sqrt{1.0+(z-\frac{R}{2})^2}} -\frac{1}{\sqrt{1.0+(z+\frac{R}{2})^2}}$ (and $\hat{W}_{ee} = 0$).
The IR pulse is taken into account using the dipole approximation and length gauge, as $\hat{v}^e_{ext}(z,t) = E(t)q_ez$,
where $E(t)=E_0\exp\left[ -\left( \frac{t-\Delta t}{\tau}\right) ^2\right]\cos(\omega (t-\Delta t))$, and
 the reduced charge $q_e=\frac{2M_{\rm H}+2}{2M_{\rm H}+1}$. The wavelength is 800 nm and the peak intensity $I_0=E_0^2=3.0\times10^{12}$W/cm$^2$. 
The pulse duration  is $\tau =4.8 fs$ and $\Delta t$ is the time delay between the UV and IR pulses. 
Here we show the results of $\Delta t=$ 7 fs.
We propagate the full TDSE~(\ref{eq:tdse})  numerically 
exactly to obtain the full molecular wavefunction
$\Psi(z,R,t)$, and from it we calculate
the probabilities of directional localization of the electron, $P_{\pm}$, 
which are defined as $P_{+(-)} = \int_{z>(<)0} dz \int dR |\Psi(z,R,t)|^2$.
These are shown as the black solid ($P_-$) and dashed ($P_+$) lines in Fig.~\ref{fig:Fig1}.
It is evident from this figure that considerable electron localization occurs, with the electron density predominantly 
localized on the left
(negative z-axis).

\begin{figure}[h]
 \centering
\includegraphics*[width=1.0\columnwidth]{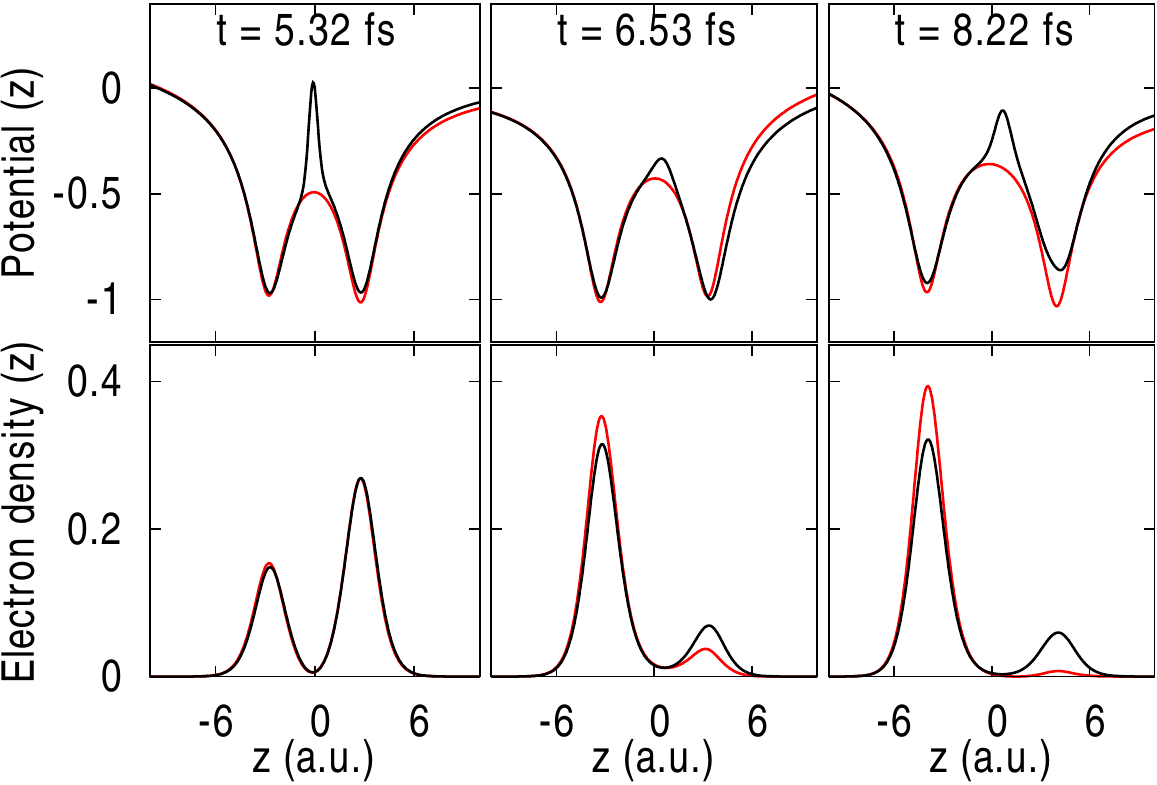}
 \caption{Top panel: Electronic potentials at the times indicated:
 (exact $\epsilon_{e}$ (black), traditional $\epsilon^{\rm trad}_{e}$
 evaluated at the exact mean nuclear position (red)).
 Lower panel: Electron densities  obtained from dynamics on the
 electronic potentials shown in the top panel. } 
  \label{fig:Fig2}
\end{figure}

We  now propagate the electrons under the traditional potential Eq.~(\ref{trad_tdpes}), 
employing the exact TD mean nuclear position
$R(t)$ obtained from $ \Psi(z,R,t)$ 
by $R(t)=\left\langle\Psi(z,R,t) \right\vert R \left\vert \Psi(z,R,t) \right\rangle$, and calculate the electron localization probabilities,
shown as red solid line (negative region) and dashed line (positive region) in Fig.~\ref{fig:Fig1}.
Comparing the red and black lines in Fig.~\ref{fig:Fig1}, we find that the traditional potential
yields the correct dynamics until around 5 fs, but then becomes less accurate: finally it predicts
the electron to be almost perfectly localized on the left nucleus,
while the exact calculation still gives some probability of finding the electron on the right.  

To understand the error in the dynamics determined by the traditional surface,
we compute the exact e-TDPES (\ref{eq:exact_etdpes}) 
in the gauge where the vector
potential $S(z,t)$ is zero~\footnote{In general, both the scalar e-TDPES~(\ref{eq:exact_etdpes}) and the vector potential~(\ref{eq:exact_evect}) 
are present. In our specific $1D$-example it is easy to see that the vector potential can be gauged away 
so that the e-TDPES remains as the 
only potential acting on the electronic subsystem.}. In the upper
panel of Fig.~\ref{fig:Fig2}, the exact $\epsilon_{e}$
(Eq. (\ref{eq:exact_etdpes})) is plotted (black line) at three
times~ \footnote{In Fig.~\ref {fig:Fig2}, curves representing $\epsilon
  _{e}$ have been rigidly shifted along the energy axis to compare with the traditional potentials},
  and compared with the traditional potential
(Eq. (\ref{trad_tdpes})) $\epsilon^{\rm trad}_{e}$ (red line) evaluated at
the exact mean nuclear position.  In the lower panel, the electron
densities calculated from dynamics on the respective potentials are
plotted.  

A notable difference between $\epsilon_{e}$ and
$\epsilon^{\rm trad}_{e}$ is an additional interatomic barrier which
appears in the exact potential, and a step-like feature that shifts
one well with respect to the other. These additional features arise
from the coupling terms contained in $\Delta\epsilon$,
and are responsible for the correct dynamics,
which is evident from
the green curve in Fig.~\ref{fig:Fig1}: this shows the results predicted by
propagating the electrons on $\epsilon^{\rm approx}_e$. The result is close to
that of the red traditional curve, and the potentials (not shown for figure clarity) are 
also close to the red potentials shown in Fig ~\ref{fig:Fig2}. A TD Hartree treatment 
is also close to the results from propagating on $\epsilon^{\rm trad}_{e}$.
An examination of the different components in Eq.~(\ref{eq:Delta_eps}) shows that 
the additional interatomic barrier arises from the term $\frac{1}{2m}\left\langle
\frac{\partial}{\partial z}\chi_z \vert \frac{\partial}{\partial z}\chi_z
\right\rangle_R$, while the  other two terms in  Eq.~(\ref{eq:Delta_eps}) yield the step.

The current understanding of the mechanism for electron localization
is that as the molecule dissociates, there is a rising interatomic
barrier from $W_{en}$,
which, when it
reaches the energy level of the excited electronic state largely shuts
off electron transfer between the ions~\cite{sansone,*HRB,*KSIV}. The
electron distribution is largely frozen after this point, as the
electron can only tunnel between the nuclei. The additional barrier we
see in the exact e-TDPES, leads to an earlier localization time, and
ultimately smaller localization asymmetry. However, each of the three
terms in Eq.~(\ref{eq:Delta_eps}) for $\Delta\epsilon$ play an important
role in the dynamics: if the electronic system is evolved adding only
the barrier correction to $\epsilon^{\rm approx}_e$ the localization
asymmetry is somewhat reduced compared to evolving on $\epsilon^{\rm
  approx}_e$ alone but far more so when all three terms of
$\Delta\epsilon$ are included.

In conclusion, we have presented the exact factorization of the
complete molecular wavefunction into electronic and nuclear
wavefunctions, $\Psi(\dulr,\dulR,t) =
\chi_\dulr(\dulR,t)\Phi(\dulr,t)$, where the electronic wavefunction
$\Phi(\dulr,t)$ satisfies 
an electronic TDSE, and the nuclear wavefunction is conditionally dependent on
the electronic coordinates.  This is complementary to the
factorization of Refs.~\cite{AMG,AMG2,AAYG}, $\Psi(\dulr,\dulR,t) =
\chi(\dulR,t)\Phi_\dulR(\dulr,t)$, where instead the nuclear
wavefunction satisfies a TDSE
while the electronic wavefunction does not.
The exact e-TDPES and exact TD vector potential acting on the electrons
were uniquely defined and compared with the traditional 
potentials used in studying localization dynamics in a model
of the H$_2^+$ molecular ion.  The importance of the exact
e-n correlation in the e-TDPES in reproducing the
correct electron dynamics was demonstrated. Further studies on this
and other model systems will lead to insight into how 
e-n correlation affects electron dynamics
in non-adiabatic processes, an insight that can never
be gained from the classical electrostatic potentials caused by
the point charges of the clamped nucleus nor the charge distributions of the exact nuclear density.
Preliminary studies using the Shin-Metiu model~\cite{Metiu} 
of field-free electronic dynamics in the presence of strong non-adiabatic couplings 
 show that peak and shift structures in the exact e-TDPES,
similar to those in the localization problem discussed here, appear typically after non-adiabatic transitions.
Finally, we note that the exact TD electronic potentials
defined in this study, together with the exact TD nuclear
potentials derived in~\cite{AMG,AMG2} establish the exact potential
functionals of TD multicomponent density functional theory~\cite{LT,KvG,KG}.  The study 
of these potentials may ultimately lead to approximate density-functionals for use in this theory, 
which holds  promise for the description 
of real-time coupled e-n dynamics in real systems.

Partial support 
from the Deutsche Forschungsgemeinschaft (SFB 762),
the European Commission (FP7-NMP-CRONOS), and the 
U.S. Department of Energy, Office of Basic Energy Sciences, Division of Chemical Sciences, 
Geosciences and Biosciences under award DE-SC0008623 (NTM),is gratefully acknowledged.

\bibliography{./etdpes}

\end{document}